\documentclass[twocolumn,english,aps,preprintnumbers,amsmath,amssymb,superscriptaddress]{revtex4}
\usepackage[latin9]{inputenc}
\usepackage{amsmath}
\usepackage{graphicx}

\makeatletter
\@ifundefined{textcolor}{}
{%
 \definecolor{BLACK}{gray}{0}
 \definecolor{WHITE}{gray}{1}
 \definecolor{RED}{rgb}{1,0,0}
 \definecolor{GREEN}{rgb}{0,1,0}
 \definecolor{BLUE}{rgb}{0,0,1}
 \definecolor{CYAN}{cmyk}{1,0,0,0}
 \definecolor{MAGENTA}{cmyk}{0,1,0,0}
 \definecolor{YELLOW}{cmyk}{0,0,1,0}
}

\usepackage{dcolumn}
\usepackage{bm}
\usepackage{color}
\usepackage[final]{pdfpages}

\makeatother

\usepackage{babel}

\begin{document}

\preprint{preprint(\today)}

\title{Ir-Sb Binary System: Unveiling Nodeless Unconventional Superconductivity Proximate to Honeycomb-Vacancy Ordering}

\author{V. Sazgari}
\affiliation{Laboratory for Muon Spin Spectroscopy, Paul Scherrer Institute, CH-5232 Villigen PSI, Switzerland}

\author{Tianping Ying}
\affiliation{Materials Research Center for Element Strategy, Tokyo Institute of Technology, Yokohama
226-8503, Japan}
\affiliation{Beijing National Laboratory for Condensed Matter Physics, Institute of Physics, Chinese Academy of Sciences, Beijing 100190, China}

\author{J.N. Graham}
\affiliation{Laboratory for Muon Spin Spectroscopy, Paul Scherrer Institute, CH-5232 Villigen PSI, Switzerland}

\author{C. Mielke III}
\affiliation{Laboratory for Muon Spin Spectroscopy, Paul Scherrer Institute, CH-5232 Villigen PSI, Switzerland}

\author{D. Das}
\affiliation{Laboratory for Muon Spin Spectroscopy, Paul Scherrer Institute, CH-5232 Villigen PSI, Switzerland}

\author{S.S.~Islam}
\affiliation{Laboratory for Muon Spin Spectroscopy, Paul Scherrer Institute, CH-5232 Villigen PSI, Switzerland}

\author{M.~Bartkowiak}
\affiliation{Laboratory for Neutron and Muon Instrumentation, Paul Scherrer Institut, CH-5232 Villigen, Switzerland}

\author{R.~Khasanov}
\affiliation{Laboratory for Muon Spin Spectroscopy, Paul Scherrer Institute, CH-5232 Villigen PSI, Switzerland}

\author{H.~Luetkens}
\affiliation{Laboratory for Muon Spin Spectroscopy, Paul Scherrer Institute, CH-5232 Villigen PSI, Switzerland}

\author{H. Hosono}
\affiliation{Materials Research Center for Element Strategy, Tokyo Institute of Technology, Yokohama
226-8503, Japan}

\author{Z.~Guguchia}
\email{zurab.guguchia@psi.ch} 
\affiliation{Laboratory for Muon Spin Spectroscopy, Paul Scherrer Institute, CH-5232 Villigen PSI, Switzerland}

\begin{abstract}

\bf{Vacancies play a crucial role in solid-state physics, but their impact on materials with strong electron-electron correlations has been underexplored. A recent study on the Ir-Sb binary system, Ir$_{16}$Sb$_{18}$ revealed a novel extended buckled-honeycomb vacancy (BHV) order \cite{YQi,TYing}. Superconductivity is induced by suppressing the BHV ordering through high-pressure growth with excess Ir atoms or isovalent Rh substitution, although the nature of superconducting pairing has remained unexplored. Here, we conduct muon spin rotation experiments probing the temperature-dependence of the effective magnetic penetration depth $\lambda_{eff}\left(T\right)$ in Ir$_{1-{\delta}}$Sb (synthesized at 5.5 GPa with $T_{\rm c}$ = 4.2 K) and ambient pressure synthesized optimally Rh-doped Ir$_{1-x}$Rh$_{x}$Sb ($x$=0.3, $T_{\rm c}$ = 2.7 K). The exponential temperature dependence of the superfluid density $n_{\rm s}$/m$^{*}$ at low temperatures indicates a fully gapped superconducting state in both samples. Notably, the ratio of $T_{\rm c}$ to the superfluid density is comparable to previously measured unconventional superconductors. A significant increase in $n_{\rm s}$/m$^{*}$ in the high-pressure synthesized sample correlates with $T_{\rm c}$, a hallmark feature of unconventional superconductivity. This correlation is intrinsic to superconductivity in the Ir-Sb binary system, with the ratio of $T_{\rm c}$ and the Fermi temperature $T_{\rm F}$ about 20 times lower than in hole-doped cuprates. We further demonstrate a similar effect induced by chemical pressure (Rh substitution) and hydrostatic pressure in Ir$_{1-x}$Rh$_{x}$Sb, highlighting that the dome-shaped phase diagram is a fundamental feature of the material. These findings underscore the unconventional nature of the observed superconductivity, and classifies IrSb as the first unconventional superconducting parent phase with ordered vacancies. We also anticipate further theoretical investigations to elucidate the microscopic relationship between superconductivity and vacancy ordering in Ir$_{16}$Sb$_{18}$.}

\end{abstract}


\maketitle

\section{Introduction}

Vacancies and defects play crucial roles in the properties of materials, and their significance is evident in various classes of materials, including transition metal dichalcogenides (TMDs) and Fe-based superconductors. In transition metal dichalcogenides like MoS$_{2}$ or WSe$_{2}$, vacancies and defects can significantly influence electronic and optical properties \cite{Ataca}. For instance, point defects such as sulfur or selenium vacancies can introduce localized states in the band gap, affecting the material's conductivity and optical absorption. Defects can serve as active sites for chemical reactions and play a role in catalysis \cite{YLuo}. Defects in TMDs can also induce magnetism and lead to interesting magnetic properties \cite{GuguchiaSciAdv,Tongay}. 

Several studies have investigated the impact of vacancy ordering in different iron-based superconductors \cite{Fang2015,Zhang2012}. For example, in iron chalcogenide superconductors such as FeSe, the ordering of selenium vacancies has been observed to influence the electronic structure and can lead to novel phenomena, including the emergence of superconductivity  \cite{TKChen}. Understanding and controlling the role of vacancies and defects in materials is therefore essential for tailoring their properties for specific applications. Researchers often explore the effects of defects to harness their potential benefits or mitigate undesirable consequences in various materials systems, ranging from electronics and catalysis to energy storage and superconductivity.

In this context, scientists have uncovered a distinctive type of vacancy ordering in the Ir-Sb binary system Ir$_{16}$Sb$_{18}$, manifesting as an extended buckled-honeycomb vacancy (BHV) order \cite{YQi,TYing}. This discovery marks a significant milestone, as Ir$_{16}$Sb$_{18}$ has been identified as the first superconducting parent phase known to exhibit ordered vacancies. The emergence of superconductivity in Ir-Sb is closely linked to the suppression of the BHV ordering. This suppression is achieved through two distinct methods: high-pressure growth of Ir$_{1-{\delta}}$Sb involving the squeezing of additional Ir atoms into the vacancies, and isovalent Rh substitution Ir$_{1-x}$Rh$_{x}$Sb. These interventions disrupt the ordered vacancy structure, paving the way for superconductivity. However, while the connection between vacancy ordering and superconductivity is established, the exact nature of the superconducting pairing in this system remains an intriguing aspect which has not yet been fully explored.

The comprehensive exploration of superconductivity at the microscopic level in the bulk of Ir$_{1-{\delta}}$Sb and optimally doped Ir$_{1-x}$Rh$_{x}$Sb ($x$ = 0.3) is essential, requiring both experimental and theoretical investigations. In this context, our focus is on muon spin rotation/relaxation/resonance ($\mu$SR) measurements of the magnetic penetration depth, $\lambda$ in these superconductors \cite{GuguchiaMoTe2,Sonier}. This parameter is fundamental to understanding superconductivity, as it is directly related to the superfluid density, $n_{s}$ through the expression, 1/${\lambda}^{2}$ = $\mu_{0}$$e^{2}$$n_{s}/m^{*}$ (where $m^{*}$ is the effective mass). The temperature dependence of ${\lambda}$ is particularly sensitive to the structure of the superconducting gap \cite{GuguchiaMoTe2,DDas}. Moreover, zero-field ${\mu}$SR proves to be a powerful tool for detecting a spontaneous magnetic field arising from time-reversal symmetry (TRS) breaking in exotic superconductors \cite{LukeTRS,HillierTRS,GuguchiaNPJ,GuguchiaMielke}. This is particularly noteworthy as internal magnetic fields as small as 0.1 G can be detected in measurements without the application of external magnetic fields. These investigations aim to unveil the intricate details of the superconducting state in Ir$_{1-{\delta}}$Sb and Ir$_{1-x}$Rh$_{x}$Sb, and contribute to our broader understanding of unconventional superconductivity in these materials.

We report on the fully gapped and time-reversal invariant superconducting state in the bulk of Ir$_{1-{\delta}}$Sb and Ir$_{0.7}$Rh$_{0.3}$Sb. The fully gapped nature suggests a well-defined energy structure in the superconducting state, while time-reversal invariance emphasizes the preservation of fundamental symmetries in the superconducting order parameter. The zero-temperature limit of the penetration depth was evaluated to be approximately X nm and Y nm for the respective materials. Notably, the $T_{\rm c}$/$\lambda_{eff}^{-2}$ ratio was found to be comparable to that of unconventional superconductors. The relatively high critical temperature ($T_{\rm c}$) despite a small carrier density raises the intriguing possibility of an unconventional pairing mechanism in Ir-Sb binary superconductors. This observation opens avenues for further investigation into the underlying physics of superconductivity in these materials, potentially revealing novel mechanisms that contribute to their unique superconducting properties.

\begin{figure}[htb!]
\includegraphics[width=1.0\linewidth]{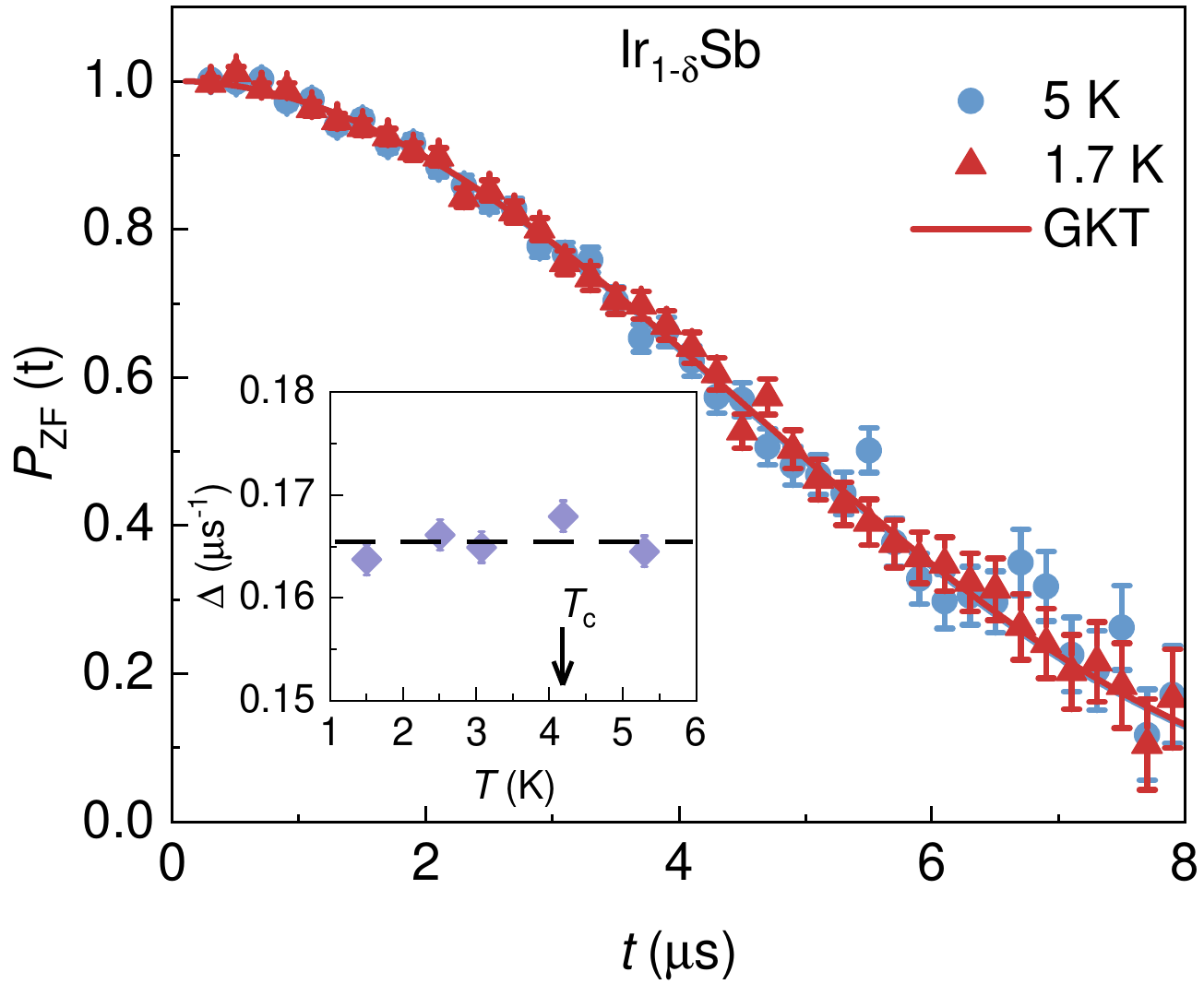}
\vspace{-0.5cm}
\caption{(Color online) \textbf{Zero-fied (ZF) ${\mu}$SR time spectra.} 
Time evolution of zero-field muon spin polarization, measured above and below $T_{\rm c}$ for Ir$_{1-{\delta}}$Sb, synthesized at 5.5 GPa. Error bars are the s.e.m. in about 10$^{6}$ events. The error of each bin count n is given by the s.d. of n. The errors of each bin in $A(t)$ are then calculated by s.e. propagation. The solid lines represent fits to the data by means of equation (1). The inset displays the temperature dependence of the zero-field muon spin relaxation rate across $T_{\rm c}$ ${\simeq}$ 4.0 K.} 
\label{fig7}
\end{figure}

\begin{figure*}[htb!]
\includegraphics[width=1.0\linewidth]{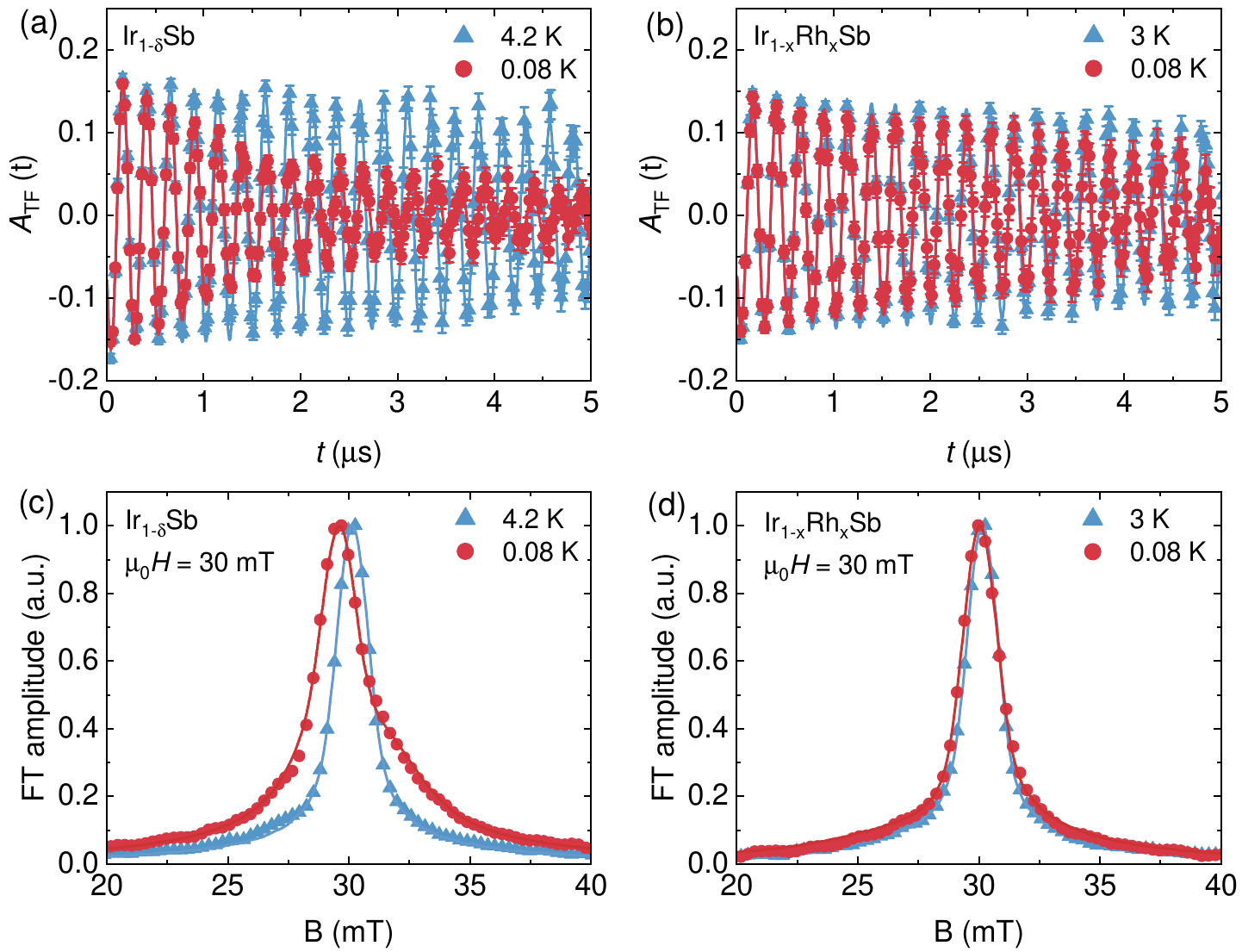}
\vspace{-0.5cm}
\caption{(Color online) \textbf{Transverse-field (TF) ${\mu}$SR time spectra and the corresponding Fourier transforms.} 
${\mu}$SR spectra are obtained above and below $T_{\rm c}$ (after field cooling the sample from above $T_{\rm c}$)
for Ir$_{1-{\delta}}$Sb, synthesized at 5.5 GPa, (a,b) and Ir$_{1-x}$Rh$_{x}$Sb with $x$ = 0.3 (c,d). Error bars are the s.e.m. in about 10$^{6}$ events. The error of each bin count n is given by the s.d. of n. The errors of each bin in $A(t)$ are then calculated by s.e. propagation. The solid lines in (a) and (c) represent fits to the data by means of equation (3). The solid lines in (b) and (d) are the Fourier transforms of the fitted time spectra.} 
\label{fig7}
\end{figure*}


\section{Results and Discussion}

The investigation into the possible magnetism, both static and fluctuating, in Ir$_{1-{\delta}}$Sb involved zero-field muon spin relaxation (ZF-${\mu}$SR) experiments conducted both above and below the critical temperature, $T_{{\rm c}}$. Figure 1 illustrates that, down to 1.5 K, no evidence of either static or fluctuating magnetism was detected in the ZF time spectra. The ZF-${\mu}$SR spectra were well-described by a damped Gaussian Kubo-Toyabe (GKT) depolarization function \cite{Toyabe}, indicative of the field distribution at the muon site generated by nuclear moments. Additionally, the absence of any change in the ZF-${\mu}$SR relaxation rate across $T_{\rm c}$ was observed, suggesting the lack of spontaneous magnetic fields associated with a time-reversal symmetry (TRS) breaking pairing state in Ir$_{1-{\delta}}$Sb. 

Figures~2a and c depict the TF-$\mu$SR time spectra for Ir$_{1-{\delta}}$Sb and Ir$_{0.7}$Rh$_{0.3}$Sb, respectively. These measurements were conducted in an applied magnetic field of 30 mT, both above (4 K) and below (0.08 K) the superconducting transition temperature $T_{\rm c}$. Above $T_{\rm c}$ the oscillations show a small relaxation due to the random local fields from the nuclear magnetic moments. At 0.08 K, the relaxation rate increases due to the formation of a flux-line lattice (FLL) in the superconducting state, resulting in a nonuniform local field distribution. It is noteworthy that the rise in relaxation rate in the superconducting state is more pronounced in Ir$_{1-{\delta}}$Sb compared to Ir$_{0.7}$Rh$_{0.3}$Sb. This distinction is further evident in the Fourier transforms (see Fig. 2b and d) of the $\mu$SR time spectra, highlighting a significant broadening of the signal in the superconducting state for Ir$_{1-{\delta}}$Sb, whilst the spectra are almost identical below $T_{\rm c}$ in  Ir$_{0.7}$Rh$_{0.3}$Sb.

\begin{figure}[t]
\includegraphics[width=1.0\linewidth]{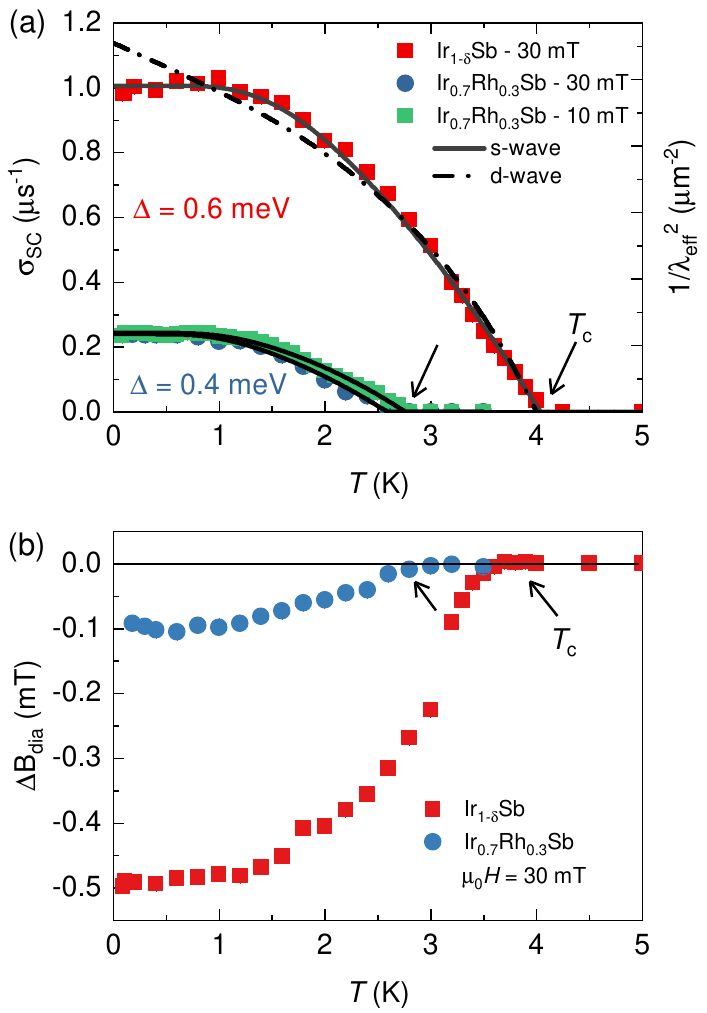}
\caption{(Color online) \textbf{Superconducting muon spin depolarization rate ${\sigma}_{\rm sc}$ and the field shift.} 
(a) Temperature dependence of the  superconducting muon spin depolarization rate, ${\sigma}_{\rm sc}$ measured in an applied magnetic fields of ${\mu}_{\rm 0}H = 10$~mT and 30 mT for Ir$_{1-{\delta}}$Sb, synthesized at 5.5 GPa and Ir$_{1-x}$Rh$_{x}$Sb with $x$ = 0.3. (b) Temperature dependence of the difference between the internal field ${\mu}_{\rm 0}$$H_{\rm SC}$ measured in the SC state and the one measured in the normal state ${\mu}_{\rm 0}$$H_{\rm NS}$ at $T$ = 5~K for Ir$_{1-{\delta}}$Sb and Ir$_{1-x}$Rh$_{x}$Sb.}
\label{fig7}
\end{figure}

As denoted by the solid lines  in Figs.~2a and c, TF-${\mu}$SR data were analyzed using the following functional form \cite{Bastian}:
\begin{equation}
A_{TF_s}(t)= \sum_{i=1}^{2}A_{s,i}e^{\Big[-\frac{(\sigma_{sc,i}^2+\sigma_{nm}^2)t^2}{2}\Big]}\cos(\gamma_{\mu}B_{int,s,i}t+\varphi).
\label{eq1}
\end{equation}
A two-component expression was utilized for the sample Ir$_{1-{\delta}}$Sb, synthesized at 5.5 GPa, owing to the asymmetric field distribution observed (see Fig. 2c). In contrast, the field distribution for the sample Ir$_{0.7}$Rh$_{0.3}$Sb is symmetric (see Fig. 2d), hence only one component was employed. In Eq. 1, $A_{s,i}$, $B_{int,s,i}$ and ${\sigma}_{\rm sc,i}$ are the the initial asymmetry, the internal magnetic field at the muon site and the superconducting relaxation rates of the i-th component. ${\sigma}_{\rm nm}$ characterizes the damping due to  the nuclear magnetic dipolar contribution. During the analysis ${\sigma}_{\rm nm}$ was assumed to be constant over the entire temperature range and was fixed to the value obtained above $T_{\rm c}$ where only nuclear magnetic moments contribute to the muon depolarization rate. In order to extract superconducting muon spin depolarization rate ${\sigma}_{\rm sc}$ (the second moment of the field distribution) and $B_{int,s}$ (the first moment of the field distribution) from the two-component fitting we used the same procedure as described in Ref. \cite{KhasanovPRL}.

In Fig.~2a, ${\sigma}_{\rm sc}$ is plotted against temperature for both Ir$_{1-{\delta}}$Sb, synthesized at 5.5 GPa (at ${\mu}_{\rm 0}H=0.03$~T), and for Ir$_{0.7}$Rh$_{0.3}$Sb (at ${\mu}_{\rm 0}H=0.01$~T and 0.03 T). Below $T_{\rm c}$, the relaxation rate ${\sigma}_{\rm sc}$ begins to increase from zero due to the formation of the flux-line lattice (FLL) and exhibits saturation at lower temperatures. The temperature dependence of ${\sigma}_{{\rm sc}}$ reflects the topology of the superconducting gap and is consistent with the presence of a single superconducting gap on the Fermi surface of these materials, as we show below. The absolute value of ${\sigma}_{\rm sc}$ is five times smaller for Ir$_{0.7}$Rh$_{0.3}$Sb compared to Ir$_{1-{\delta}}$Sb, indicating a lower superfluid density for the Rh-doped sample. Below $T_{\rm c}$, a large diamagnetic shift of $B_{\text{int,s}}$ experienced by the muons is observed in both samples. In Fig. 2b, the temperature dependence of the diamagnetic shift ${\delta}B_{dia}$ = $B_{\text{int,s,SC}}$ - $B_{\text{int,s,NS}}$ is plotted, where $B_{\text{int,s,SC}}$ represents the internal field measured in the superconducting state, and $B_{\text{int,s,NS}}$ is the internal field measured in the normal state at 5 K. This diamagnetic shift indicates the bulk nature of superconductivity and rules out the possibility of field-induced magnetism in these superconductors.

To perform a quantitative analysis, it is important to note that the London magnetic penetration depth $\lambda(T)$ is directly related related to the measured relaxation rate in the superconducting state $\sigma_{\rm sc}$. For triangular FLL relationship is described by the equation \cite{Brandt}:
\begin{equation}
\frac{\sigma_{sc}^2(T)}{\gamma_\mu^2}=0.00371\frac{\Phi_0^2}{\lambda^4(T)},
\end{equation}
where ${\Phi}_{\rm 0}=2.068 {\times} 10^{-15}$~Wb is the magnetic-flux quantum. Equation (2) is applicable only when the separation between vortices is smaller than ${\lambda}$. In this particular scenario, as per the London model, ${\sigma}_{\rm sc}$ becomes field-independent \cite{Brandt}.

To explore the superconducting gap structure of Ir$_{1-{\delta}}$Sb and Ir$_{0.7}$Rh$_{0.3}$Sb, we conducted an analysis of the temperature dependence of the magnetic penetration depth, ${\lambda}(T)$, directly linked to the superconducting gap. The behavior of ${\lambda}$($T$) can be characterized within the local (London) approximation (${\lambda}$ ${\gg}$ ${\xi}$) using the following expression \cite{Bastian,Tinkham}:
\begin{equation}
\frac{\lambda^{-2}(T,\Delta_{0,i})}{\lambda^{-2}(0,\Delta_{0,i})}=
1+\frac{1}{\pi}\int_{0}^{2\pi}\int_{\Delta(_{T,\varphi})}^{\infty}(\frac{\partial f}{\partial E})\frac{EdEd\varphi}{\sqrt{E^2-\Delta_i(T,\varphi)^2}},
\end{equation}
where $f=[1+\exp(E/k_{\rm B}T)]^{-1}$ is the Fermi function, ${\varphi}$ is the angle along the Fermi surface, and ${\Delta}_{i}(T,{\varphi})={\Delta}_{0,i}{\Gamma}(T/T_{\rm c})g({\varphi}$) (${\Delta}_{0,i}$ is the maximum gap value at $T=0$). The temperature dependence of the gap is approximated by the expression ${\Gamma}(T/T_{\rm c})=\tanh{\{}1.82[1.018(T_{\rm c}/T-1)]^{0.51}{\}}$,\cite{carrington} while $g({\varphi}$) describes the angular dependence of the gap and it is replaced by 1 for both an $s$-wave gap, [1+acos(4${\varphi}$))/(1+a)] for an anisotropic $s$-wave gap and ${\mid}\cos(2{\varphi}){\mid}$ for a $d$-wave gap \cite{Fang}.


In Fig~3a, the experimentally obtained $\lambda_{eff}^{-2}$($T$) dependence is most accurately described by a momentum-independent $s$-wave model with a gap value of $\Delta$ = 0.6(1)~meV and $T_{\rm c}$ = 4.1(1)~K for Ir$_{1-{\delta}}$Sb, and a gap value of $\Delta$ = 0.4(1)~meV and $T_{\rm c}$ = 2.7(2)K for Ir$_{0.7}$Rh$_{0.3}$Sb. The $d$-wave and $p$-wave gap symmetries were also considered but were found to be inconsistent with the data (illustrated by the dashed line in Fig. 3a). Particularly, these models struggle to account for the very weak temperature dependence of ${\lambda}$($T$) at low temperatures. Additionally, the power-law behavior 
$\left[1-\left(\frac{T}{T_c}\right)^2\right]$, theoretically proposed for the superfluid density of dirty $d$-wave superconductors \cite{Hirshfeld}, was tested but deemed inconsistent with the data. This analysis shows that a nodeless or fully gapped state is the most plausible bulk superconducting pairing state for Ir$_{1-{\delta}}$Sb and Ir$_{0.7}$Rh$_{0.3}$Sb.

The estimated ratio of the superconducting gap to $T_{\rm c}$, (2$\Delta/k_{\rm B}T_{\rm c}$), is approximately 3.4, aligning with the BCS (Bardeen-Cooper-Schrieffer) expectation \cite{GuguchiaMoTe2}. However, it's crucial to acknowledge that a similar ratio can also be anticipated within a Bose-Einstein Condensation (BEC)-like framework. Importantly, the ratio 2$\Delta/k_{\rm B}T_{\rm c}$, on its own, does not effectively distinguish between BCS or BEC condensation scenarios. Further insights are required to differentiate between these two possibilities and elucidate the nature of the superconducting state in the studied materials. What distinguishes between BCS and BEC superconductivity is a key parameter: the ratio of the superconducting critical temperature to the superfluid density. This ratio, $T_{\rm c}/n_s$, plays a crucial role in characterizing the nature of the superconducting state in different materials. In a simplified interpretation of the BEC to BCS crossover, the $T_{\rm c}/n_s$ ratio serves as a critical parameter. Systems characterized by a small $T_{\rm c}/n_s$ (with a large superfluid density, $n_s$) are often considered to reside on the "BCS" side of the crossover. Conversely, systems with a large $T_{\rm c}/n_s$ (exhibiting a small superfluid density, $n_s$) are expected to be on the BEC side. Moreover, the correlation between $T_{\rm c}$ and the superfluid density is anticipated to be significant primarily on the BEC side of the crossover. As one moves from the BCS limit to the BEC limit in the crossover, the nature of the pairing mechanism evolves, transitioning from Cooper pairs formed through electron-phonon interactions (BCS) to a more Bose-Einstein condensation-like scenario involving preformed pairs. 



\begin{figure}[t!]
\includegraphics[width=1.0\linewidth]{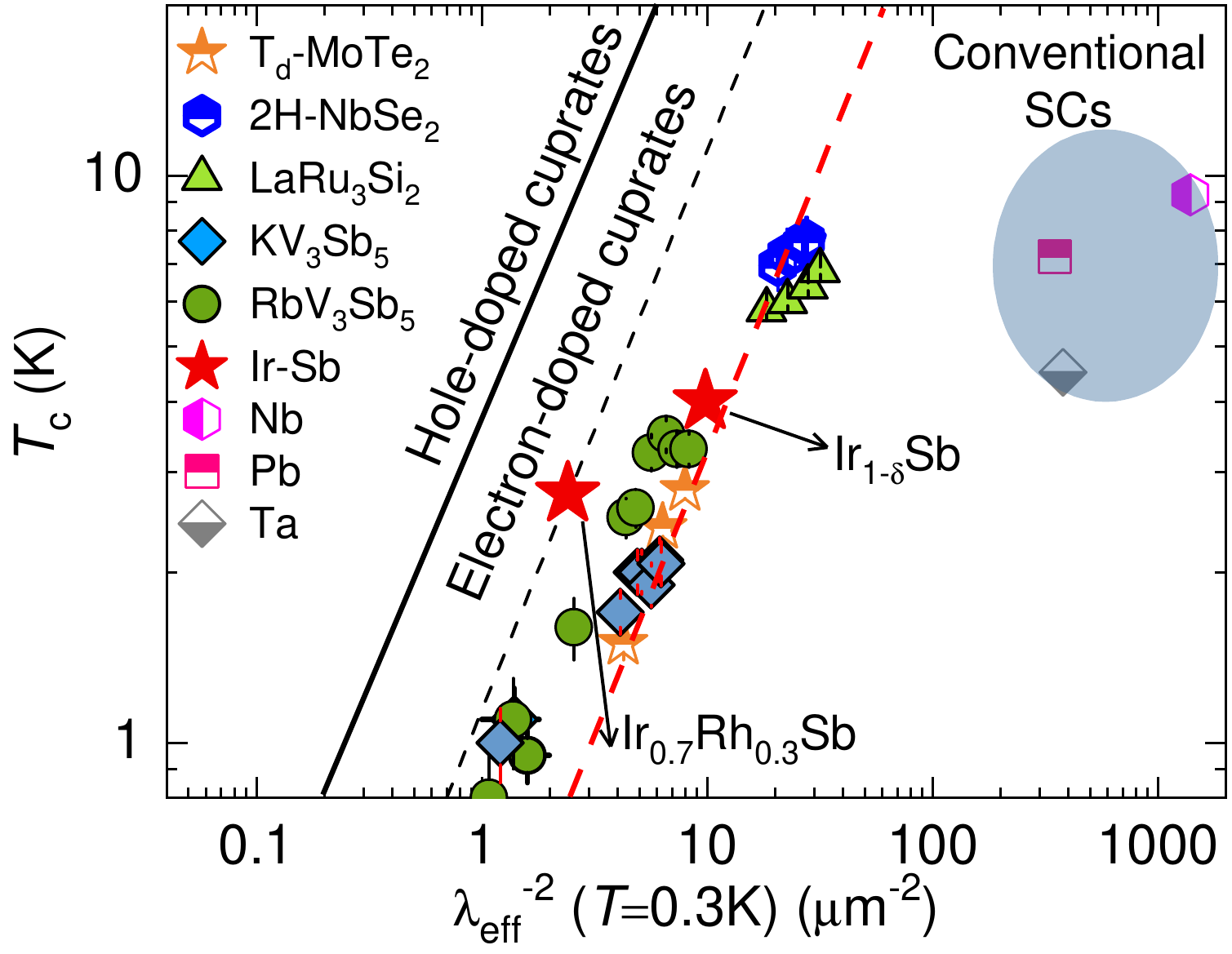}
\caption{(Color online) \textbf{Hallmark feature of unconventional superconductivity.} 
Plot of $T_{\rm c}$ versus ${\lambda}_{\rm eff}^{-2}(0)$ on a logarithmic scale obtained from ${\mu}$SR experiments for Ir$_{1-{\delta}}$Sb, synthesized at 5.5 GPa and Ir$_{0.7}$Rh$_{0.3}$Sb. The data for the kagome-lattice superconductors KV$_{3}$Sb$_{5}$ \cite{GuguchiaMielke,GuguchiaRVS,GuguchiaNPJ}, RbV$_{3}$Sb$_{5}$ \cite{GuguchiaRVS}, and LaRu$_{3}$Si$_{2}$ \cite{GuguchiaPRM} are also included. The dashed red line represents the relationship obtained for the layered transition metal dichalcogenide superconductors, $T_{d}$-MoTe$_{2}$ and 2H-NbSe$_{2}$ by Guguchia $et~al$. \cite{GuguchiaNbSe2,GuguchiaNature}. The relationship observed for cuprates is shown \cite{Uemura1,Shengelaya} as well as the points for various conventional superconductors.}
\label{fig7}
\end{figure}

\begin{figure*}[htb!]
\includegraphics[width=1.0\linewidth]{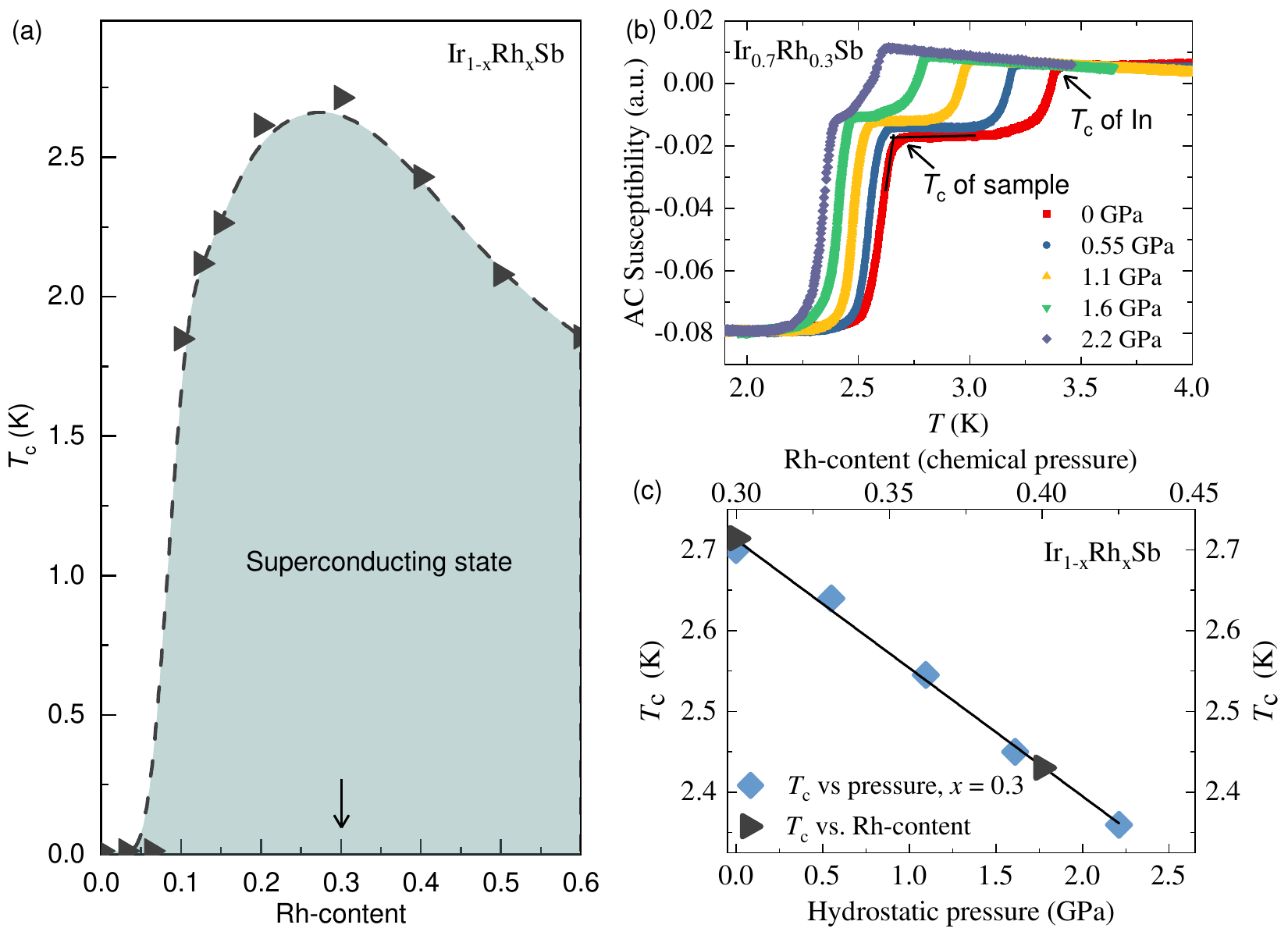}
\vspace{-0.5cm}
\caption{(Color online) \textbf{Chemical and hydrostatic pressure effects on superconductivity in Ir$_{1-x}$Rh$_{x}$Sb.} (a) The superconducting critical temperature vs Rh-content, showing a dome-shaped SC phase diagram (adapted from Ref. \cite{YQi}). Arrows indicate the sample with $x$ = 0.3, which was measured under hydrostatic pressure. (b) The temperature dependence of AC susceptibility, measured at various hydrostatic pressure, ranging up to 2.21 GPa. (c) The superconducting critical temperature as a function of hydrostatic pressure and the Rh-content ("chemical pressure") in the range between $x$ = 0.3 and 0.45.}
\label{fig7}
\end{figure*}

The observation of a correlation between $T_{\rm c}$ and the superfluid density ($\lambda_{eff}^{-2}$) was first noted in hole-doped cuprates back in 1988-89 \cite{Uemura1,Uemura2}, extending later to include electron-doped cuprates \cite{Shengelaya}. This intriguing relationship has been investigated across various superconducting systems. Guguchia and collaborators demonstrated that this linear correlation is an intrinsic feature in superconductors such as transition metal dichalcogenides \cite{GuguchiaMoTe2,GuguchiaNbSe2} and kagome-lattice superconductors \cite{GuguchiaNature,GuguchiaRVS,GuguchiaNPJ}. The ratio $T_{\rm c}$/$\lambda_{eff}^{-2}$ in these systems tends to be lower than that observed in hole-doped cuprates  (see Figure 4). To contextualize the superconductors Ir$_{1-{\delta}}$Sb and Ir$_{0.7}$Rh$_{0.3}$Sb within this framework, Fig. 4 illustrates the critical temperature plotted against the superfluid density. For Ir$_{0.7}$Rh$_{0.3}$Sb, the estimated ratio $T_{\rm c}$/$\lambda_{eff}^{-2}$ is approximately 1, closely resembling electron-doped cuprates known for their correlated superconductivity. In the case of Ir$_{1-{\delta}}$Sb, the ratio is reduced to 0.4 but remains notably distant from conventional BCS superconductors. Intriguingly, it aligns nearly perfectly with the trend line occupied by charge density wave superconductors like 2H-NbSe$_{2}$, 4H-NbSe$_{2}$, LaRu$_{3}$Si$_{2}$, as well as the Weyl-superconductor $T_{d}$-MoTe$_{2}$ \cite{GuguchiaMoTe2}. This finding strongly suggests an unconventional pairing mechanism in Ir$_{1-{\delta}}$Sb and Ir$_{0.7}$Rh$_{0.3}$Sb, characterized by a low density of Cooper pairs.

Another unconventional feature in the superconducting phase diagram of Ir$_{1-x}$Rh$_{x}$Sb is a dome-shaped dependence of $T_{\rm c}$ (see Figure 5a). This pattern is characterized by an optimal $T_{\rm c}$ value occurring at $x$ = 0.3, followed by a reduction as the Rh concentration deviates from this optimal point. The isovalent Rh substitution on the Ir site in IrSb, without introducing additional holes or electrons, creates a condition often termed "chemical pressure". Typically, chemical substitution introduces disorder effects, potentially influencing $T_{\rm c}$. To discern the intrinsic nature of this dome shape, a cleaner external parameter is essential. For example, hydrostatic pressure introduces fewer disorder effects compared to isovalent chemical substitutions ("chemical pressure"). For this reason, we explored the impact on $T_{\rm c}$ in the optimally Rh-doped, Ir$_{0.7}$Rh$_{0.3}$Sb with hydrostatic pressure, spanning a range up to $p = 2.2$ GPa. As shown in Figs. 5b and c, the observed trend revealed a linear decrease in $T_{\rm c}$ with increasing pressure. This behavior aligns closely with the effects of Rh doping. This finding suggests that the impact of both external pressure conditions applied hydrostatically and induced through chemical modifications is consistent in the Ir$_{0.7}$Rh$_{0.3}$Sb system. This highlights the intrinsic nature of the reduction in $T_{\rm c}$ beyond $x$ = 0.3 and enhance our understanding of the unique aspects of superconductivity in this material.


\section{CONCLUSIONS}

In summary, our study provides a microscopic exploration of superconductivity in Ir$_{1-{\delta}}$Sb (synthesized at 5.5 GPa with $T_{\rm c}$ = 4.2 K) and optimally Rh-doped Ir$_{0.7}$Rh$_{0.3}$Sb ($T_{\rm c}$ = 2.7 K) in close proximity to vacancy ordering, employing a bulk sensitive local probe. Specifically, we investigated the zero-temperature magnetic penetration depth ${\lambda}_{eff}\left(0\right)$ and the temperature dependence of ${\lambda_{eff}^{-2}}$ through ${\mu}$SR experiments. The superfluid density in both systems aligns with a scenario of a complete gap. Intriguingly, the $T_{\rm c}$/$\lambda_{eff}^{-2}$ ratio is comparable to that of high-temperature unconventional superconductors, suggesting the unconventional nature of superconductivity in Ir-Sb binary superconductors. Additionally, the ${\mu}$SR experiments, serving as an extremely sensitive magnetic probe, do not exhibit evidence of spontaneous magnetic fields, which would be expected for a time-reversal-symmetry-breaking state in the bulk of the superconductor. Consequently, our results categorize Ir$_{1-{\delta}}$Sb and Ir$_{0.7}$Rh$_{0.3}$Sb as unconventional, time-reversal-invariant, and fully gapped bulk superconductors. We further demonstrate the striking similarity between the effects of chemical pressure, induced by isovalent Rh substitution, and hydrostatic pressure on the superconducting critical temperature in Ir$_{0.7}$Rh$_{0.3}$Sb. This highlights that the observed dome-shaped dependence of $T_{\rm c}$ is not merely a consequence of disorder effects introduced by chemical substitution but is rooted in the intrinsic properties of the material. These results offer valuable insights into the underlying mechanisms governing the material's behavior. A more comprehensive analysis requires consideration of various factors, including the specific pairing mechanisms and the role of interactions in the superconducting state.




\section{Acknowledgments}~
The ${\mu}$SR experiments were carried out at the Swiss Muon Source (S${\mu}$S) Paul Scherrer Insitute, Villigen, Switzerland. Z.G. acknowledges support from the Swiss National Science Foundation (SNSF) through SNSF Starting Grant (No. TMSGI2${\_}$211750). Z.G. acknowledges the useful discussions with Robert Scheuermann. T.Y. would like to acknowledge Beijing Natural Science Foundation (Grant No. Z200005).\\

\textbf{Author Contributions:}  
Z. Guguchia conceived and supervised the project. Sample Growth: Y.P. and H.H.. $\mu$SR experiments, data analysis and corresponding discussions: V.S., J.N.G.,  D.D., C.M.III., S.S.I., R.K., H.L., and Z.G.. Figure development and writing of the paper: Z.G., and V.S., with contributions from all authors. All authors discussed the results, interpretation and conclusion.\\

\textbf{Competing interests:} The authors declare that they have no competing interests.\\

\section{Methods}

\textbf{Sample growth}: The details of the synthesis of the polycrystalline samples of Ir$_{1-{\delta}}$Sb and Ir$_{0.7}$Rh$_{0.3}$Sb are reported elsewhere \cite{YQi,TYing}.\\

\textbf{Experimental details}: Zero field (ZF) and transverse field (TF) $\mu$SR experiments  were performed on the GPS (${\pi}$M3 beamline) \cite{Amato}, and high-field HAL-9500 instruments (${\pi}$E3 beamline) \cite{Sedlak}, equipped with BlueFors vacuum-loaded cryogen-free dilution refrigerator (DR), at the Swiss Muon Source (S$\mu$S) at the Paul Scherrer Institute, in Villigen, Switzerland. Zero field is dynamically obtained (compensation better than 30 mG) by a newly installed automatic compensation device \cite{Amato}. When performing measurements in zero-field the geomagnetic field or any stray fields are tabulated and automatically compensated by the automatic compensation device. The ${\mu}$SR time spectra were analyzed using the free software package MUSRFIT \cite{Bastian}. AC susceptibility experiments under pressure were carried out using double wall piston-cylinder type of cell made of MP35N/MP35N material generate pressures up to 2.3 GPa \cite{GuguchiaPressure,GuguchiaNature,Khasanov2022}. A small indium plate was placed together with the sample in the pressure cell filled with the Daphne oil. The pressure was estimated by tracking the SC transition of a indium plate by AC susceptibility.\\

\end{document}